\documentclass[pdflatex,sn-mathphys-num]{sn-jnl}
\usepackage{graphicx}%
\usepackage{multirow}%
\usepackage{amsmath,amssymb,amsfonts}%
\usepackage{amsthm}%
\usepackage{mathrsfs}%
\usepackage[title]{appendix}%
\usepackage{xcolor}%
\usepackage{textcomp}%
\usepackage{manyfoot}%
\usepackage{booktabs}%
\usepackage{algorithm}%
\usepackage{algorithmicx}%
\usepackage{algpseudocode}%
\usepackage{listings}%

\begin{document}

\title[Article Title]{Bayesian quantum phase estimation with fixed photon states}

\author[1]{\fnm{Boyu} \sur{Zhou}}\email{zhouboyu@arizona.edu}

\author[2,3]{\fnm{Saikat} \sur{Guha}}\email{saikat@umd.edu}

\author*[4,3,5]{\fnm{Christos N.} \sur{Gagatsos}}\email{cgagatsos@arizona.edu}

\affil[1]{\orgdiv{Department of Physics}, \orgname{The University of Arizona}, \orgaddress{\street{1118 E 4th St}, \city{Tucson}, \postcode{85721}, \state{Arizona}, \country{USA}}}

\affil[2]{\orgdiv{Department of Electrical and Computer Engineering}, \orgname{ University of Maryland}, \orgaddress{\street{2410 A.V. Williams Building}, \city{College Park}, \postcode{20742}, \state{Maryland}, \country{USA}}}

\affil[3]{\orgdiv{Wyant College of Optical Sciences}, \orgname{The University of Arizona}, \orgaddress{\street{1630 E University Blvd}, \city{Tucson}, \postcode{85721}, \state{Arizona}, \country{USA}}}

\affil*[4]{\orgdiv{Department of Electrical and Computer Engineering}, \orgname{The University of Arizona}, \orgaddress{\street{1230 E Speedway Blvd}, \city{Tucson}, \postcode{85721}, \state{Arizona}, \country{USA}}}

\affil[5]{\orgdiv{Program in Applied Mathematics}, \orgname{The University of Arizona}, \orgaddress{\street{617 N. Santa Rita}, \city{Tucson}, \postcode{85721}, \state{Arizona}, \country{USA}}}

\abstract{We consider a two-mode bosonic state with fixed photon number $n \in \mathbb{N}$, whose upper and lower modes pick up a phase $\phi$ and $-\phi$ respectively. We compute the optimal Fock coefficients of the input state, such that the mean square error (MSE) for estimating $\phi$ is minimized while the minimum MSE is always attainable by a measurement. Our setting is Bayesian, i.e., we consider $\phi$ to be a random variable that follows a prior probability distribution function (PDF). Initially, we consider the flat prior PDF and we discuss the well-known fact that the MSE is not an informative tool for estimating a phase when the variance of the prior PDF is large. Therefore, we move on to study truncated versions of the flat prior in both single-shot and adaptive approaches. For our adaptive technique we consider $n=1$ and truncated prior PDFs. Each subsequent step utilizes as prior PDF the posterior probability of the previous step and at the same time we update the optimal state and optimal measurement.}



\maketitle

\section{Introduction}\label{sec1}

Quantum phase sensing is a field of study that has attracted the attention of the scientific community because of its high relevance to quantum technologies and its theoretical importance in studying quantum-enhanced sensing protocols by using non-classical phenomena such as squeezing and entanglement, which improve the scaling of the quantum Fisher information (QFI) \cite{Lane1993,Giovannetti2011,Pezze2015,Crowley2014,Hump2013,Lang2013}.

The vast majority of the literature elaborates on and expands the knowledge of the Fisherian approach, i.e., when the estimated parameter has a fixed, yet unknown, value. In this approach, typically one engages with evaluating lower bounds on the mean squared error of an estimator. Said bounds are the well-known quantum and classical Cram\'er-Rao bounds, given by the inverse of the QFI and its classical counterpart, the classical Fisher information (CFI), aiming to show a quantum enhancement of the sensing performance and to reveal the measurement that attains the quantum bound. Another direction to consider is the Bayesian setting: In this approach, the estimated parameter is considered as a random value that follows some probability distribution function (PDF), which is called the \emph{prior} PDF. For single-variable Bayesian sensing tasks, exact formulas have been given for evaluating the minimum mean square error (MMSE), instead of lower bounding the MSE, in contrast to the Fisherian approach where the QFI serves as a lower bound to the Fisherian MSE. Said Bayesian MMSE is guaranteed to be always attainable by a measurement. In this paper we consider estimation of a single variable in the Bayesian approach, and specifically we consider the MMSE, not the Bayesian versions of the Cram\'er-Rao bound 
\cite{VanTrees2013,Gill1995}, on which notable works such as \cite{Jarzyna2015,Morelli_2021} have explored further. Moreover, it is worthwhile to note that multi-variable Bayesian lower bounds on the covariance matrix of an estimator have been established \cite{Rubio2020,Sidhu2020}, while the conditions of their attainability are well-understood \cite{Rubio2020,Sidhu2020}. Other notable works on precision bounds (Bayesian and more general than the Cram\'er-Rao bounds) include \cite{Mankei2012,Lu2016,Rubio_2018,Hall_2012,Teklu2009}.

Even though the mathematical setting for the Bayesian MMSE approach is known, compared to the QFI approach there are very few works exploiting and expanding it, some examples are \cite{Rubio_2019,Rubio2020,rubio2024}. As we note later in the paper and as already has been observed \cite{zhou2023,Li2018}, there is no direct correspondence between the Fisherian and the Bayesian approaches, as the two approaches assign different meaning to the concept of probability \cite{Li2018}. Therefore, studying the quantum Bayesian sensing (QBS) is beneficial for two main reasons: It can give new results that may not be predicted by a Fisherian approach (or any Bayesian bounds build upon their Fisherian counterparts), which in turn will help to develop new intuition on Bayesian sensing, and it is a natural setting for adaptive techniques \cite{Morelli_2021,Rubio_2018,Rubio_2019,Lee2022,Wiebe2016,Brivio2010} and special focus has been given to utilizing Bayesian methods to update the controlling parameter of a fixed measurement \cite{berry2000optimal,higgins2007entanglement,berry2009perform,xiang2011entanglement,wiseman1997adaptive,smith2023adaptive,martinez2019adaptive}. In this paper, we consider a single phase as our unknown variable, therefore our analytic and numerical calculations pertain to the actual MMSE, not to a lower bound on the MSE.

The structure and main results of this work are: In Section \ref{sec:setting} we explain our physical setting and we briefly give the mathematical tools we utilize. In Section \ref{sec:MMSEeval} we compute the MMSE for the generic state with fixed photon number and for a flat prior PDF and we discuss why we proceed with truncated, step-like, prior PDFs. Then, we consider an important special case, i.e., a NOON state. As we discuss (and has already been pointed out before \cite{Demkowicz2011}), the MSE is a fundamentally problematic metric when the prior PDF has large variance, such as the flat PDF. This is reflected in the results we derive pertaining to the NOON state and that is why we repeat the analysis for truncated versions of the flat prior PDF. In Section \ref{sec:adaptive} we fix the photon number to be $n=1$ and we give a fully-optimized example of an adaptive phase sensing technique based on the Bayesian approach: Starting with a different truncated flat prior PDFs and their corresponding optimal state for each prior PDF, for each step we update the prior PDF, the optimal input state, and the optimal measurement. Finally, in Section \ref{sec:concl} we discuss our results and further research directions.

\section{The setting}\label{sec:setting}
Let the estimated single variable to be $\phi \in [0,2\pi]$, which in our case represents an optical phase shifting. The Bayesian MSE is defined as \cite{Personick1971},
\begin{eqnarray}
  \label{eq:deltaB}  \delta_B = \int_0^{2\pi} d\phi P(\phi) \text{tr}\left[\hat{\rho}(\phi)(\hat{H}-\phi\hat{I})^2\right],
\end{eqnarray}
where $\hat{\rho}(\phi)$ is the state that holds the parameter (i.e. the final state), $P(\phi)$ is the prior PDF, and $\hat{H}$ is a Hermitian operator whose eigenvectors represent the measurement that we utilize to acquire information on $\phi$.

We note that the MSE as defined in Eq. \eqref{eq:deltaB} is equivalent to the (perhaps) most commonly used form,
\begin{eqnarray}
    \delta_B = \int_0^{2\pi} d\phi \sum_k P(\phi) P(k|\phi) (h_k-\phi)^2,
    \label{eq:deltaB2}
\end{eqnarray}
where $P(k|\phi)=\langle h_k |\hat{\rho}(\phi) |h_k\rangle $, $|h_k\rangle$ and $h_k$ are respectively the eigenvectors and eigenvalues of $\hat{H}$. If the spectrum of $\hat{H}$ is continuous, the index $k$ is replaced by $y$ and the summation over $k$ by an integral over $y$. The equivalence of Eqs. \eqref{eq:deltaB} and \eqref{eq:deltaB2} can be shown be starting from Eq. \eqref{eq:deltaB} and writing the operator $\hat{H}$ in its diagonal form.

Using Eq. \eqref{eq:deltaB} it has been shown \cite{Personick1971} that minimizing over all possible $\hat{H}$, i.e., by solving the functional minimization problem,
\begin{eqnarray}
\delta \equiv \underset{\hat{H}}{\text{min}}\ \delta_B,
\end{eqnarray}
we get the MMSE,
\begin{eqnarray}
\label{eq:delta1}\delta = \text{tr}\hat{\Gamma}_2 - \text{tr}(\hat{B} \hat{\Gamma}_1),
\end{eqnarray}
where,
\begin{eqnarray}
  \label{eq:Gammak}  \hat{\Gamma}_k = \int_0^{2\pi} d\phi P(\phi) \phi^k \hat{\rho}(\phi),\ k=0,1,2,
\end{eqnarray}
and
\begin{eqnarray}
\label{eq:B}\hat{B} = 2 \int_0^\infty dze^{-z \hat{\Gamma}_0} \hat{\Gamma}_1 e^{-z \hat{\Gamma}_0},
\end{eqnarray}
where the MMSE $\delta$ is always attainable by the optimal projective measurement which is given by the eigenvectors of $\hat{B}$. We note that Eqs. \eqref{eq:deltaB} and \eqref{eq:deltaB2} are not the most general expressions for the MSE as they assume protective measurements. However, it has been shown that by starting with general positive operator valued measures, optimality is not lost by considering just projective measurements \cite[Appendix A therein]{Macieszczak2014}.

The operator $\hat{B}$ is a solution to the equation $\hat{B}\hat{\Gamma}_0+\hat{\Gamma}_0\hat{B}=2 \hat{\Gamma}_1$ \cite{Personick1971}. However, said equation admits other possible solutions of the form $\hat{B}'=\hat{B}+\hat{K}$, under the conditions $\hat{K}\hat{\Gamma}_0=\hat{\Gamma}_0\hat{K}=0$ and $\hat{K}=\hat{K}^\dagger$. The eigenvectors of any operator $\hat{B}'$ provide an optimal projective measurement while $\delta$ remains the same. This is proven in Appendix \ref{app:Bprime} by following the same reasoning as in the QFI-based work \cite{Shi_2023}.

In this work, we consider states whose Fock basis expansion has the form,
\begin{eqnarray}
    |\Psi_n\rangle = \sum_{l=0}^n a_l |l,n-l\rangle.
    \label{eq:GenericState}
\end{eqnarray}
The mean photon number of such states is always $\bar{n}=n$ regardless of the values of the complex coefficients $a_l$ which satisfy,
\begin{eqnarray}
    \sum_{l=0}^n |a_l|^2 =1.
    \label{eq:normalization}
\end{eqnarray}
Since the mean photon number satisfies $\bar{n}=n$, the only condition the probe state of Eq. \eqref{eq:GenericState} must respect is the normalization given in Eq. \eqref{eq:normalization}.

The phase sensing performance, within the Fisherian framework, of states in the form of Eq. \eqref{eq:GenericState} has been examined in \cite{Dorner2009} where it was proven to be significantly improved compared to classical interferometers.  
\begin{figure}[H]
\centering
\includegraphics[width=0.9\textwidth]{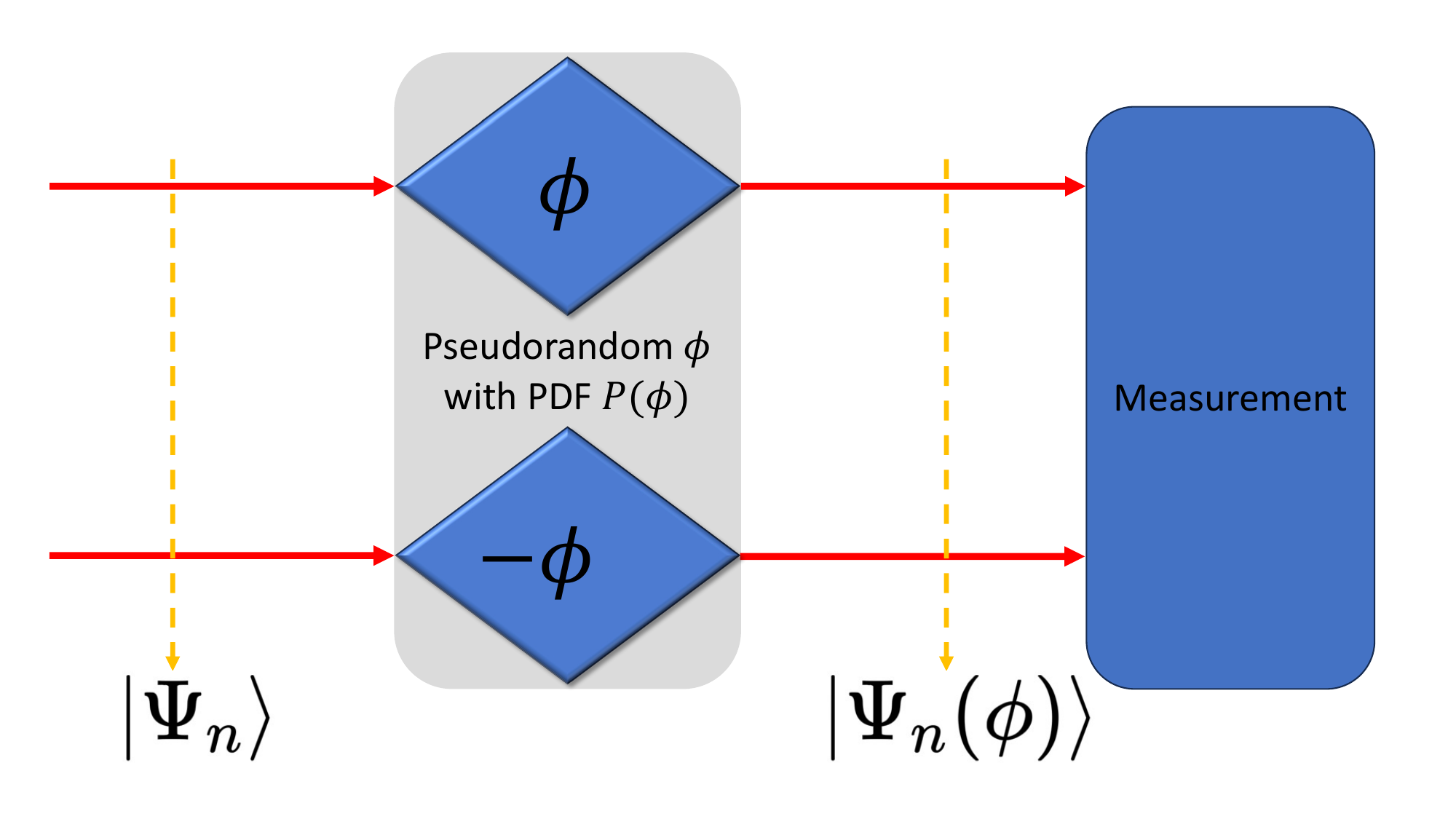}
\caption{The upper and lower modes of the state of Eq. \eqref{eq:GenericStatePhi} are phase shifted by $\phi$ and $-\phi$ respectively, resulting to the state of Eq. \eqref{eq:GenericStatePhi}. The phase is a pseudorandom number which follows a PDF $P(\phi)$, i.e., what we refer to in the text as prior PDF. The phase shifting is modelled by the unitary operator $\hat{U}(\phi)=e^{i \phi(\hat{n}_1-\hat{n}_2)}$, where $\hat{n}_1$ and $\hat{n}_2$ are the number operators defined on the Hilbert spaces of the upper and lower mode respectively.}
\label{fig:Phase}
\end{figure}
In our setting (see Fig. \ref{fig:Phase}), the upper and lower modes of the input state $|\Psi_n\rangle$ pick up a phase $\phi$ and $-\phi$ respectively, resulting to the output state,
\begin{eqnarray}
    |\Psi_n (\phi)\rangle = \sum_{l=0}^n a_l e^{i (2 l - n) \phi} |l,n-l\rangle.
    \label{eq:GenericStatePhi}
\end{eqnarray}

\section{MMSE evaluations}\label{sec:MMSEeval}
\subsection{The generic state and the flat prior PDF}\label{sec:generic}
Our immediate tasks entail to computing $\delta$ of Eq. \eqref{eq:delta1} for $\hat{\rho}(\phi)= |\Psi_n (\phi)\rangle \langle \Psi_n (\phi) |$ and for the flat prior PDF,
\begin{eqnarray}
    P(\phi)=\frac{1}{2\pi},
    \label{eq:flatPrior}
\end{eqnarray}
which represents minimal prior knowledge \cite{Demkowicz2010,Demkowicz2011}, as opposed to a Dirac delta prior PDF $P_\delta(\phi)=\delta(\phi-\phi_0)$ which renders the MMSE equal to zero, i.e., full information on $\phi$ has already been given in the form of prior PDF. The choice of the flat prior can be justified due to translational invariance. The flat prior PDF is also a natural assumption to initiate an adaptive technique, i.e., since we know nothing on the parameter a priori, we assume that all values have the same probability to occur. Lastly, it can be the choice of an adversary who holds the phase and can manipulate it in a way that does not favor some values over others.

It is important to note that the choice of the MSE as the cost function using a large variance prior PDF, like the one of Eq. \eqref{eq:flatPrior}, is far from being optimal. Indeed, let us say that at a given instance the true value of the phase is $0 \leq \phi_0 \ll 1$ while the estimated phase is $\tilde{\phi}=2\pi-\epsilon$, then the MSE will register a much larger error than the true one, simply because the MSE is not periodic in $\phi$. 
Therefore, it appears proper to employ a different error metric, i.e., a periodic one. However, while progress has been made towards a general framework minimizing periodic metrics analytically \cite{rubio2024,Demkowicz2011}, numerical evalutions are in many cases unavoidable (even though such a research directions is interesting in its own means, see for example \cite{Bavaresco2024}). In any case, the choice of the MSE as the cost function ceases to be problematic when the the variance of the prior PDF has a small variance \cite{Rubio2020}. 
The goal of this section is to partly show the shortcomings of the MSE when using the flat prior PDF by evaluating Eq. \eqref{eq:delta1} for the state of Eq. \eqref{eq:GenericStatePhi} and the flat prior PDF of Eq. \eqref{eq:flatPrior}. From Eq. \eqref{eq:Gammak} we find,
\begin{eqnarray}
    \hat{\Gamma}_k = \sum_{l,l'=0}^{n} a_l a_{l'}^* b_k^{(l-l')} |l,n-l\rangle \langle l',n-l'|, 
    \label{eq:Gammak2}
\end{eqnarray}
where,
\begin{eqnarray}
    b_k^{(l-l')} = \frac{1}{2\pi} \int_0^{2\pi} d\phi \phi^k e^{2 i \phi (l-l') },
    \label{eq:bllp}
\end{eqnarray}
which for $k=0,1,2$ is evaluated to,
\begin{eqnarray}
 \label{eq:b0}   b_0^{(l-l')} &=& \delta_{l,l'},\\
 \label{eq:b1}   b_1^{(l-l')} &=& \left\{ \begin{array}{ll}
     \pi, & l=l'   \\
     -\frac{i}{2(l-l')}, & l\neq l'
 \end{array}\right.,\\
  \label{eq:b2}   b_2^{(l-l')} &=& \left\{ \begin{array}{ll}
     \frac{4\pi^2}{3}, & l=l'   \\
     \frac{1-2 i \pi (l-l')}{2(l-l')^2}, & l\neq l'
 \end{array}\right.,
\end{eqnarray}
where $\delta_{l,l'}$ denotes the Kronecker delta.

We note that the lower branch of Eq. \eqref{eq:b2} is not necessary since we are interested only in the diagonal terms of $\hat{\Gamma}_2$ (per Eq. \eqref{eq:delta1} we only need the trace of $\hat{\Gamma}_2$). In any case we have calculated said lower branch for cross-checking reasons and for completeness. In fact we find,
\begin{eqnarray}
    \label{eq:TrGamma2} \text{tr}\hat{\Gamma}_2 = \frac{4\pi^2}{3}.
\end{eqnarray}
From Eqs. \eqref{eq:Gammak2} and \eqref{eq:b0} we find,
\begin{eqnarray}
    \label{eq:Gamma0} \hat{\Gamma}_0 = \sum_{l=0}^n |a_l|^2 |l,n-l\rangle \langle l,n-l|,
\end{eqnarray}
from which, since $\hat{\Gamma}_0$ is diagonal on the Fock basis, we readily find,
\begin{eqnarray}
 \label{eq:expGamma0}   e^{-z \hat{\Gamma}_0} = \sum_{l=0}^n e^{-z |a_l|^2} |l,n-l\rangle \langle l,n-l|.
\end{eqnarray}
From Eqs. \eqref{eq:B}, \eqref{eq:Gammak2} (for $k=1$), \eqref{eq:b1}, \eqref{eq:expGamma0}, and by paying attention that $e^{-z \hat{\Gamma}_0}$ and $\hat{\Gamma}_1$ do not commute, the second term of Eq. \eqref{eq:delta1} gives,
\begin{eqnarray}
\label{eq:trBGamma1}    \text{tr}(\hat{B}\hat{\Gamma}_1) = \pi^2 + \sum_{l,l'=0 \atop l\neq l'}^n \frac{|a_l|^2 |a_{l'}|^2}{2(l-l')^2(|a_l|^2+|a_{l'}|^2)}.
\end{eqnarray}
Finally, from Eqs. \eqref{eq:delta1}, \eqref{eq:TrGamma2} and \eqref{eq:trBGamma1} we find the MMSE,
\begin{eqnarray}
    \delta = \frac{\pi^2}{3}-\sum_{l,l'=0 \atop l\neq l'}^n \frac{|a_l|^2 |a_{l'}|^2}{2(l-l')^2(|a_l|^2+|a_{l'}|^2)}.
    \label{eq:deltaGeneric}
\end{eqnarray}
We note that the MMSE depends only the modulo of the complex coefficients $a_l = |a_l|e^{i \theta_l}$. The choice of the flat prior PDF surely played a role in that; as we will see in Section \ref{sec:adaptive}, the choice of other prior PDFs does not result to such feature.

The optimal projective measurement is given by the eigenvectors of $\hat{B}$, a calculation that in principle is doable (analytically or numerically) since the matrix representation of said operator is finite. In Section \ref{sec:adaptive} we do an example for $n=1$.

\subsection{The NOON state and the flat prior PDF}\label{sec:noonFlat}
The NOON state,
\begin{eqnarray}
    |\Psi_{\text{NOON}}\rangle = \frac{1}{\sqrt{2}} \left(|n,0\rangle+|0,n\rangle\right),
    \label{eq:noonState}
\end{eqnarray}
where $n\geq 1,\ n\in \mathbb{N}$, is a special case of the state given in Eq. \eqref{eq:GenericState}, i.e., for $a_{l}=1/\sqrt{2}(\delta_{l,0}+\delta_{l,n})$, Eq. \eqref{eq:GenericState} gives Eq. \eqref{eq:noonState}. NOON states are of central importance in quantum information theory and specifically in quantum-enhanced sensing \cite{Dorner2009,Grun2022b}, while their implementation has been suggested recently \cite{Grun2022a}.

We now consider the NOON state as the input state of Fig. \ref{fig:Phase} and the flat prior PDF of Eq. \eqref{eq:flatPrior}. Applying Eq. \eqref{eq:deltaGeneric} we get the MMSE,
\begin{eqnarray}
    \delta_{\text{NOON}} = \frac{\pi^2}{3}-\frac{1}{4n^2}.
    \label{eq:deltaNOON}
\end{eqnarray}
The MMSE of Eq. \eqref{eq:deltaNOON}, albeit positive $\forall\ n\geq1$ (it attains its minimal value for $n=1$), is a manifestation of the problematic use of the MSE when a large-variance prior is considered. It is also a manifestation of the fundamental technical difference between the Bayesian and Fisherian approaches: How their mathematical formulations use the information of the prior PDF. Equation \eqref{eq:deltaNOON} increases monotonically with $n$, while its Fisherian counterpart, i.e., the QFI $F$ (since we consider a single parameter, the QFI can be equal to a CFI \cite{Braunstein1994}), scales as $n^2$ \cite{Dorner2009}, i.e., $F^{-1} \propto n^{-2}$, with a \emph{positive} coefficient, in contrast to the negative coefficient of $n^{-2}$ in Eq. \eqref{eq:deltaNOON}. We note that any QFI-based lower bound on the Bayesian MSE \cite{VanTrees2013}, will not have the qualitative behavior of Eq. \eqref{eq:deltaNOON} simply because the Fisher information is a positive quantity. 

We note that since for the flat prior PDF, Eq. \eqref{eq:deltaGeneric} depends only on the absolute values of the coefficients $a_l$, the MMSE of Eq. \eqref{eq:deltaNOON} will remain invariant even if we assume that the non-zero Fock coefficients of the NOON state have an imaginary part, while their absolute value is always equal to $1/\sqrt{2}$.
We plot Eq. \eqref{eq:deltaNOON} as function of $n$ in Fig. \ref{fig:deltaNOON}.

In \cite{Rubio2020}, the equivalent of Eq. \eqref{eq:deltaNOON}, behaves as $n^2$ because the flat prior PDF therein depends on $n$ (the larger the $n$, the smaller the prior PDF's variance). Therefore, as discussed in Section \ref{sec:setting}, the seemingly counter-intuitive behavior of Eq. \eqref{eq:deltaNOON} is attributed to the large variance of the prior PDF, and the disregard to the modulo-$2\pi$ wraparound of optical phase by the MMSE metric. 

\subsection{The NOON state and the truncated flat prior PDF}\label{sec:noonTFlat}
We now consider truncated versions of the flat prior and the NOON state. We consider as input state the NOON state of Eq. \eqref{eq:noonState} and the truncated version of the flat prior as the prior PDF,
\begin{eqnarray}
  \label{eq:truncated PDF} 
  P_{\text{trunc}}(\phi)=\frac{1}{m}, \phi \in (0,m],
  \label{eq:flatPriorTrun}
\end{eqnarray}
where the parameter $m \in (0,2\pi]$ controls the size of the truncated length, i.e., for $m=2 \pi$, Eq. \eqref{eq:flatPriorTrun} gives Eq. \eqref{eq:flatPrior}. Using Eq. \eqref{eq:Gammak}, we get,
\begin{eqnarray}
  \label{eq:truncated Gamma_k} 
 \hat{\Gamma}_k^{(\text{trunc})} = \begin{pmatrix}
 d_{11} & d_{12} \\
 d_{12}^* & d_{11} \\
\end{pmatrix},
\end{eqnarray}
where,
\begin{eqnarray}
    d_{11}&=&\frac{m^k}{2 k+2},\\
    d_{12}&=&\frac{2^{-k-2} (i n)^{-k-1} (\gamma (k+1)-\gamma (k+1,2 i m n))}{m},
\end{eqnarray}
and $\gamma (\cdot)$ is the incomplete Gamma function. From Eqs. \eqref{eq:delta1}, \eqref{eq:B} and \eqref{eq:truncated Gamma_k} we get the MMSE for the NOON input state and the truncated flat prior,
\begin{eqnarray}
  \label{eq:truncated MMSE} 
\delta_{\text{trunc}}=\frac{2 m^4 n^4-3 m^2 n^2+\left(3-3 m^2 n^2\right) \cos (2 m n)+6 m n \sin (2 m n)-3}{24 m^2 n^4}.
\end{eqnarray}
We note that for $m=2 \pi$, since $n$ is always an integer, $\sin (2 \pi n)=0$ and $\cos (2 \pi n)=1$. Therefore, for said case Eq. \eqref{eq:truncated MMSE} gives Eq. \eqref{eq:deltaNOON}.

\begin{figure}[h]
\centering
\includegraphics[width=0.9\textwidth]{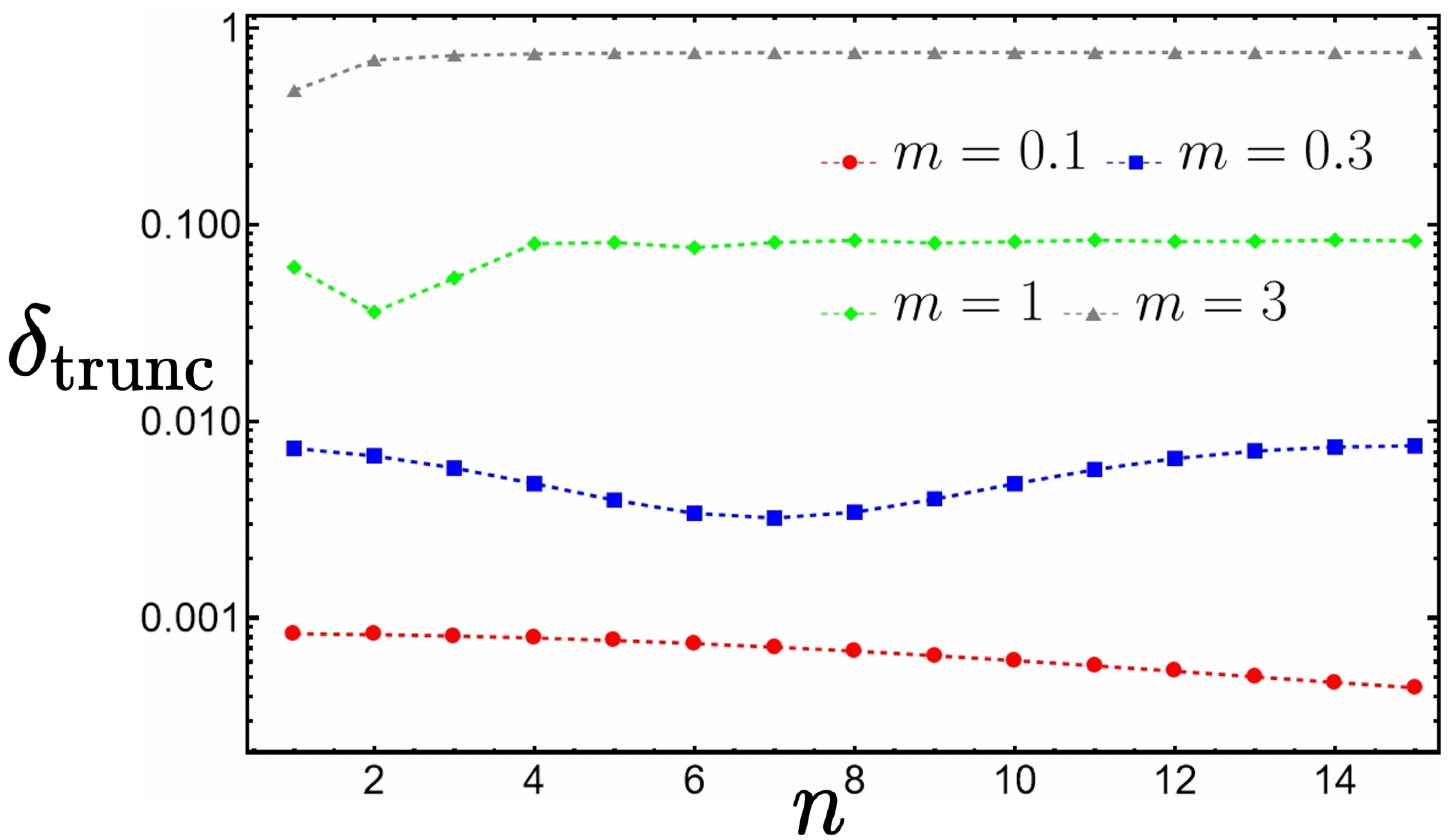}
\caption{The MMSE $\delta_{\text{trunc}}$ for the truncated flat prior PDF for NOON state, as a function of the photon number $n$. The y-axis is in logarithmic scale. The red circle is $m=0.1$, the blue square is $m=0.3$, the green diamond is $m=1$, and the gray triangle is $m=3$.}
\label{fig:truncated MMSE}
\end{figure}

The role of parameter $m$ is shown in Fig. \ref{fig:truncated MMSE}. The MMSE decreases with $m$. This is because when $m$ decreases, we eliminate the range of $\phi$, i.e., the prior PDF becomes more informative. Also, the smaller the $m$, the minimum value $\delta_{\text{trunc}}$ as function of $n$ moves to the left. For $m=0.1$, although we can only see the downtrend in Fig. \ref{fig:truncated MMSE}, it can verified analytically from Eq. \eqref{eq:truncated MMSE}) that it starts increasing at around $n=20$. Similarly, for smaller values of $m$. Displacing the minimum of the MMSE to higher $n$ means that the smaller the $m$, the more meaningful the MSE cost functions becomes due to the smaller variance of the prior PDF. 

We close this section by noting that fixed photon states (including the NOON state) using Personick's Bayesian formulation have been studied in \cite{Branford_2021}, however the prior PDFs used are different than those presented in this work.

\section{Adaptive technique}\label{sec:adaptive}
\subsection{Preliminaries and description}
For simplicity, we will use the notation $|\bar{0}\rangle = |01\rangle$ and $|\bar{1}\rangle = |10\rangle$, where $\{|\bar{0}\rangle,|\bar{1}\rangle\}$ is the dual-rail qubit basis. Therefore, we write the states of Eqs. \eqref{eq:GenericState} and \eqref{eq:GenericStatePhi} respectively for $n=1$ as,
\begin{eqnarray}
 \label{eq:InState_i}   |\Psi_1\rangle  &=& a_0|\bar{0}\rangle+a_1|\bar{1}\rangle,\\
\label{eq:OutState_i} |\Psi_1 (\phi)\rangle  &=& a_0 e^{-i \phi}|\bar{0}\rangle+a_1 e^{i \phi}|\bar{1}\rangle.
\end{eqnarray}
The two-mode phase shifting operation of Fig. \ref{fig:Phase} now reads $e^{-i \phi\hat{\sigma}_z}$, where $\hat{\sigma}_z$ is the Pauli $z$ operator such as $\hat{\sigma}_z |\bar{0}\rangle = |\bar{0}\rangle $ and $\hat{\sigma}_z |\bar{1}\rangle = -|\bar{1}\rangle $.

All matrix and vector representations in this section are understood in the dual-rail qubit basis. The optimal projective measurement is given by the eigenvectors of the operator $\hat{B}$. Since we consider a phase estimation problem and we fix the input state to be of the form of Eq. \eqref{eq:InState_i}, the dimensions of the matrix representation of $\hat{B}$ on the dual-rail qubit basis, will always be $2 \times 2$ for everything that we will discuss in this section (and Appendix \ref{app:adaptiveFlat}). Therefore, any possible measurement will have two possible outcomes.

Now let us discuss how we go from one step to the next. For each step $s$, we denote the quantities of interest, i.e., the prior PDF, the optimal state, the operator $\hat{B}$ (whose eigenvectors provide the optimal projective measurement), and the optimal (minimum) MMSE respectively as follows, $P^{(s;j_1\ldots j_{s-1})}(\phi)$, $|\Psi_1^{(s;j_1\ldots j_{s-1})}\rangle$, $\hat{B}^{(s;j_1\ldots j_{s-1})}$, $\delta^{(s;j_1\ldots j_{s-1})}$. We also denote the eigenvectors of $\hat{B}^{(s;j_1\ldots j_{s-1})}$ as $|v_1^{(s;j_1\ldots j_{s-1})}\rangle$ and $|v_2^{(s;j_1\ldots j_{s-1})}\rangle$, while the corresponding eigenvalues are $v_1^{(s;j_1\ldots j_{s-1})}$ and $v_2^{(s;j_1\ldots j_{s-1})}$. The indices $j_i$ take values $j_i=1,2\ \forall i=1,\ldots,s-1$.

For the first step ($s=1$), we define $(1;j_1\ldots j_{0}) \equiv (1)$, rendering the optimal state equal to $|\Psi_1^{(s)}\rangle \equiv |\Psi_1^{\text{opt}}\rangle$, i.e., the state of Eq. \eqref{eq:InState_i} with the optimal coefficients. Then we find the matrix representation in the dual-rail basis of $\hat{B}$ and we calculate the probability of each possible outcome for said optimal measurement and finally the optimal MMSE. The connection between any two sequential steps is given by Bayes rule,
  \begin{eqnarray}
    P(\phi|v_{j_{s-1}}^{(s-1;j_1\ldots j_{s-2})})= \frac{P(v_{j_{s-1}}^{(s-1;j_1\ldots j_{s-2})}|\phi) P^{(s-1;j_1\ldots j_{s-2})}(\phi)}{\int_0^{2\pi} d\phi P(v_{j_{s-1}}^{(s-1;j_1\ldots j_{s-2})}|\phi) P^{(s-1;j_1\ldots j_{s-2})}(\phi)}
    \label{eq:BayesRule}
\end{eqnarray}  
and utilizing the conditional probability $P(\phi|v_{j_{s-1}}^{(s-1;j_1\ldots j_{s-2})})$ as the prior PDF of the subsequent step, i.e.,
\begin{eqnarray}
    P^{(s;j_1\ldots j_{s-1})}(\phi) \equiv P(\phi|v_{j_{s-1}}^{(s-1;j_1\ldots j_{s-2})}).
\end{eqnarray}
Therefore, to find the prior PDF for step $s$, we must compute the probability $P(v_{j_{s-1}}^{(s-1;j_1\ldots j_{s-2})}|\phi)$ and use the prior PDF $P^{(s-1;j_1\ldots j_{s-2})}(\phi)$ of the previous step. To compute $P(v_{j_{s}}^{(s;j_1\ldots j_{s-1})}|\phi)$ (of step $s$), we use Born's rule which requires to find the eigenvectors of $\hat{B}^{(s;j_1\ldots j_{s-1})}$. 

\begin{figure}
\centering
\includegraphics[width=0.9\textwidth]{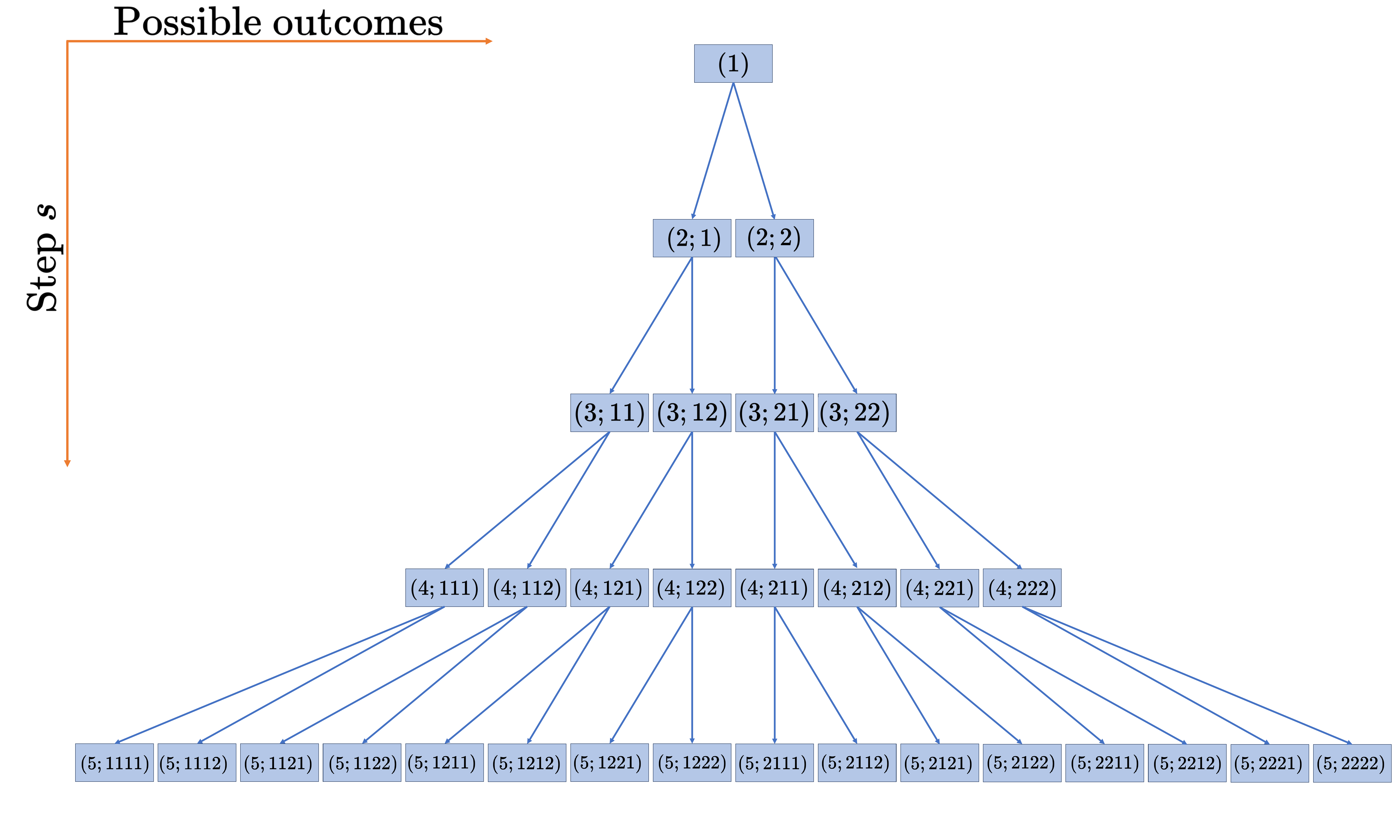}
\caption{The quantities in the boxes represent the superscripts $(s;j_1\ldots j_{s-1})$ for each step $s$ of the quantities of interest: The prior PDF $P^{(s;j_1\ldots j_{s-1})}(\phi)$, the optimal state  $|\Psi_1^{(s;j_1\ldots j_{s-1})}\rangle$, the operator $\hat{B}^{(s;j_1\ldots j_{s-1})}$, and the optimal MMSE $\delta^{(s;j_1\ldots j_{s-1})}$. For the step $s=1$, we define $(1;j_1\ldots j_{0}) \equiv (1)$. The arrows represent the two possible measurement outcomes $v_1^{(s;j_1\ldots j_{s-1})}$ and $v_2^{(s;j_1\ldots j_{s-1})}$. If one can continues growing the tree-like diagram, the number of leaves (i.e. the blue boxes) is $2^{s-1}$ for step $s$.}
\label{fig:Adaptive2}
\end{figure}

For step $s$ we have $2^{s-1}$ possible $\hat{B}^{(s;j_1\ldots j_{s-1})}$ operators. To understand why, we consider the following example: Let $s=1$, then we have one $\hat{B}$ operator, that has two eigenvectors. Each one of the possible outcomes will correspond to a different probability, leading to a different prior PDF for the next step $s=2$. Using these two prior PDF for step $s=2$, we find two different $\hat{B}$ operators. Then in step $s=3$, each one of the aforesaid two $\hat{B}$ operators will produce two possible measurement outcomes, i.e., four possible outcomes in total, and consequently four possible prior PDFs for the next step. For the same reason, for step $s$ we have in general $2^{s-1}$ different optimal states and $2^{s-1}$ optimal MMSEs, which are calculated using Eqs. \eqref{eq:delta1}, \eqref{eq:Gammak} and \eqref{eq:B}. The adaptive protocol is summarized in Fig. \ref{fig:Adaptive2}.

In what follows (including Appendix \ref{app:adaptiveFlat}), we have examined (mostly numerically) $s=5$ steps and all possible \begin{eqnarray}
\nonumber && (\text{prior PDF for}\ s=1)+(\text{prior PDFs for}\ s=2)+(\text{prior PDFs for}\ s=3)\\
\nonumber && +(\text{prior PDFs for}\ s=4)+(\text{prior PDF for}\ s=5)\\
\nonumber && =2^0+2^1+2^2+2^3+2^4=31  
\end{eqnarray} total possible quantities of interest. 

\subsection{Truncated flat prior PDF}\label{sec:AdaptiveTrunFlat}
In this section we consider truncated priors because they render the MSE a proper metric when they are narrow enough. Specifically, we start with the prior PDFs $P(\phi)=\frac{1}{\pi}$, $\phi \in [0,\pi]$,
$P(\phi)=\frac{2}{\pi}$, $\phi \in [0,\frac{\pi}{2}]$, and $P(\phi)=\frac{10}{\pi}$, $\phi \in [0,\frac{\pi}
{10}]$. Then we optimize numerically over the input fixed photon states for given $n$ and we evaluate the minimum MMSE for each case. The optimal coefficients $a_l^{\text{opt}}$ of the generic fixed photon number input state of Eq. \eqref{eq:GenericState} and the minimum MMSE are shown in Tables \ref{tab:OptVals_PI}, \ref{tab:OptVals_05PI}, and \ref{tab:OptVals_01PI}. In said tables, we observe that optimal coefficients $a_l^{\text{opt}}$ are the same regardless of the choice of the truncated prior PDF.

\begin{table}[h]
\caption{The prior PDF is $P(\phi)=\frac{1}{\pi}$, $\phi \in [0,\pi]$. Values of the optimal coefficients $a_l^{\text{opt}}$ for $n\leq 5$ and the corresponding MMSE $\delta^{\text{opt}}_{n}$.}
\label{tab:OptVals_PI}
\begin{tabular}{cccccccc}
\toprule
 $n$ & $\delta^{\text{opt}}_{n}$ & $a_0$ & $a_1$ & $a_2$ & $a_3$ & $a_4$ & $a_5$  \\
\hline
1 & 0.572467 & 0.707107 & 0.707107 & / & / & / & / \\ \hline
2 & 0.44203 & 0.55108 & 0.626595 & 0.55108 & / & / & / \\ \hline
3 & 0.361202 & 0.453382 & 0.542627 & 0.542627 & 0.453382 &/ &/ \\ \hline
4 & 0.305933 & 0.386101 &0.474686 &0.501197 &0.474686 &0.386101 &/ \\ \hline
5 & 0.265637 & 0.336767 &0.420815 &0.457715 &0.457715 &0.420815 &0.336767 \\
\botrule
\end{tabular}
\end{table}

\begin{table}[h]
\caption{The prior PDF is $P(\phi)=\frac{2}{\pi}$, $\phi \in [0,\frac{\pi}{2}]$. Values of the optimal coefficients $a_l^{\text{opt}}$ for $n\leq 5$ and the corresponding MMSE $\delta^{\text{opt}}_{n}$.}
\label{tab:OptVals_05PI}
\begin{tabular}{cccccccc}
\toprule
 $n$ & $\delta^{\text{opt}}_{n}$ & $a_0$ & $a_1$ & $a_2$ & $a_3$ & $a_4$ & $a_5$  \\
\hline
1 & 0.104296 & 0.707107 & 0.707107 & / & / & / & / \\ \hline
2 & 0.0664533 & 0.55108 & 0.626595 & 0.55108 & / & / & / \\ \hline
3 & 0.0468982 & 0.453382 & 0.542627 & 0.542627 & 0.453382 &/ &/ \\ \hline
4 & 0.0352759 & 0.386101 &0.474686 &0.501197 &0.474686 &0.386101 &/ \\ \hline
5 & 0.0276983 & 0.336767 &0.420815 &0.457715 &0.457715 &0.420815 &0.336767 \\
\botrule
\end{tabular}
\end{table}

\begin{table}[h]
\caption{The prior PDF is $P(\phi)=\frac{10}{\pi}$, $\phi \in [0,\frac{\pi}{10}]$. The values of the optimal coefficients $a_l^{\text{opt}}$ for values of $n\leq 5$ and the corresponding MMSE $\delta^{\text{opt}}_{n}$.}
\label{tab:OptVals_01PI}
\begin{tabular}{cccccccc}
\toprule
 $n$ & $\delta^{\text{opt}}_{n}$ & $a_0$ & $a_1$ & $a_2$ & $a_3$ & $a_4$ & $a_5$  \\
\hline
1 & 0.00795939 & 0.707107 & 0.707107 & / & / & / & / \\ \hline
2 & 0.00760144 & 0.55108 & 0.626595 & 0.55108 & / & / & / \\ \hline
3 & 0.00717076 & 0.453382 & 0.542627 & 0.542627 & 0.453382 &/ &/ \\ \hline
4 & 0.00669102 & 0.386101 &0.474686 &0.501197 &0.474686 &0.386101 &/ \\ \hline
5 & 0.0061858 & 0.336767 &0.420815 &0.457715 &0.457715 &0.420815 &0.336767 \\
\botrule
\end{tabular}
\end{table}

We then proceed to compute the evolution of the prior PDFs per adaptive step and we then compute the optimal single photon state, the optimal measurement and the corresponding MMSE. We see that as we progress with the steps of the adaptive technique, the prior PDFs become narrower and they have a single global maximum, while the MMSE trends to $0$.

Then, we compare the two strategies: (i) \emph{Adaptive method} with $1$-photon states and (ii) \emph{single-state-use method} (i.e. non-adaptive) with an optimized generic state whose photon number is the same as the step number of the adaptive method. For strategy (ii), the prior PDF is not updated and it is fixed to be a truncated flat prior.

\begin{figure}[H]
\centering
\includegraphics[width=0.9\textwidth]{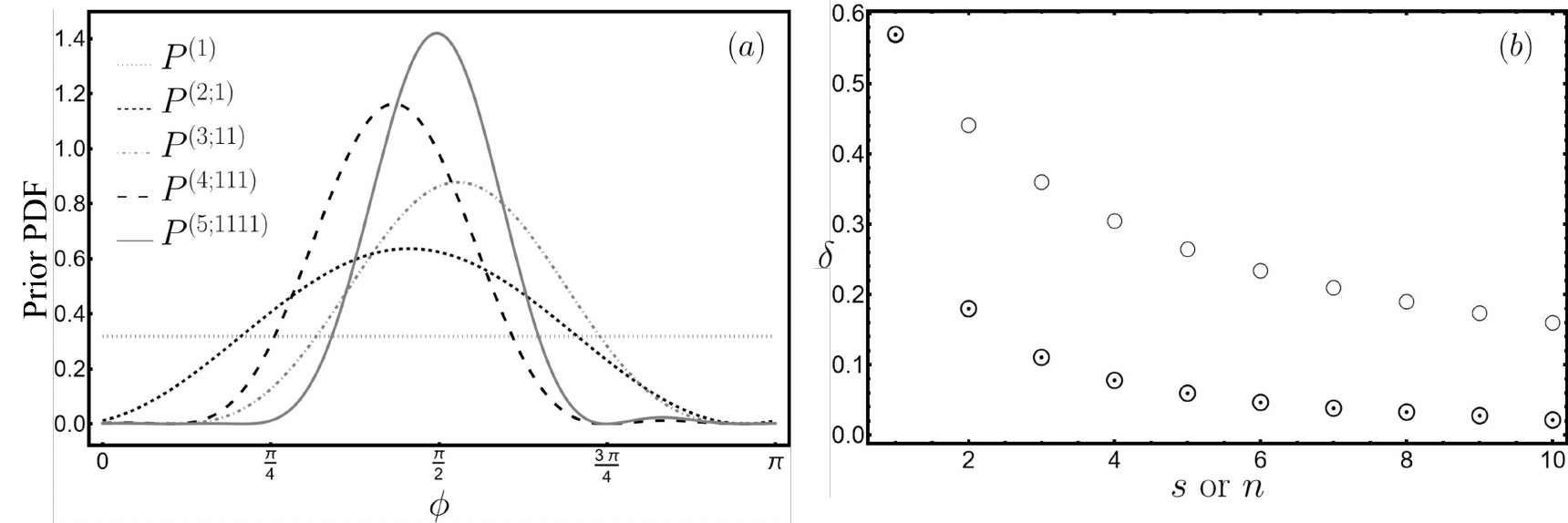}
\caption{The prior PDF is $P(\phi)=\frac{1}{\pi}$, $\phi \in [0,\pi]$. (a) The evolution of the prior PDFs for the leftmost path of Fig. \ref{fig:Adaptive2}. (b) The MMSEs of the adaptive technique (circles with dots) and the single-state-use technique (empty circles).}
\label{fig:adaptive_PI}
\end{figure}

\begin{figure}[H]
\centering
\includegraphics[width=0.9\textwidth]{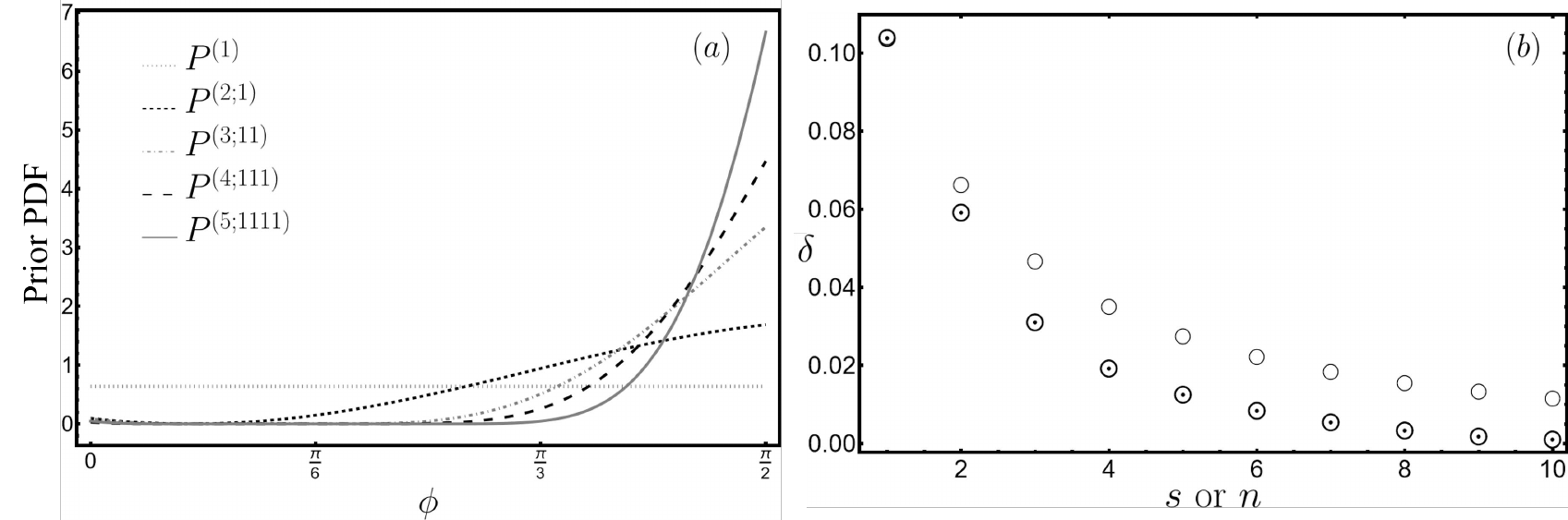}
\caption{The prior PDF is $P(\phi)=\frac{2}{\pi}$, $\phi \in [0,\frac{\pi}{2}]$. (a) The evolution of the prior PDFs for the leftmost path of Fig. \ref{fig:Adaptive2}. (b) The MMSEs of the adaptive technique (circles with dots) and the single-state-use technique (empty circles).}
\label{fig:adaptive_05PI}
\end{figure}

\begin{figure}[H]
\centering
\includegraphics[width=0.9\textwidth]{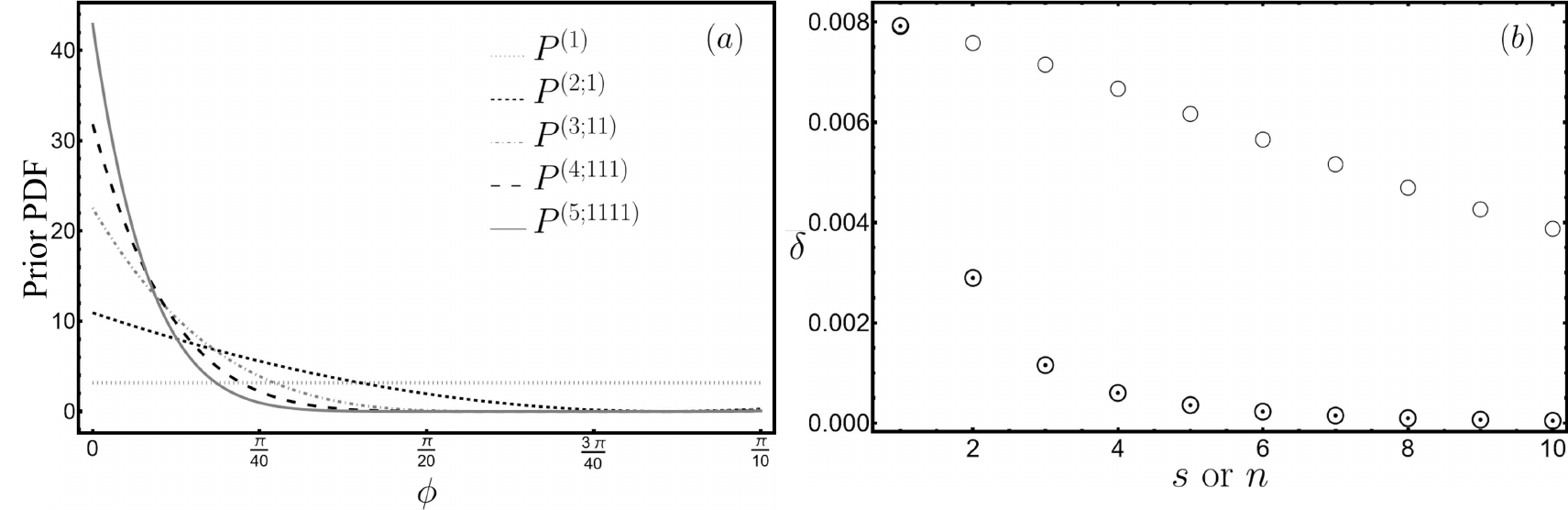}
\caption{The prior PDF is $P(\phi)=\frac{10}{\pi}$, $\phi \in [0\frac{\pi}{10}]$. (a) The evolution of the prior PDFs for the leftmost path of Fig. \ref{fig:Adaptive2}. (b) The MMSEs of the adaptive technique (circles with dots) and the single-state-use technique (empty circles).}
\label{fig:adaptive_01PI}
\end{figure}

This comparison is fair in the sense that both strategies use the same number of photons either one by one per step or all in a single sensing attempt. We show our results in Figs. \ref{fig:adaptive_PI}, \ref{fig:adaptive_05PI}, and \ref{fig:adaptive_01PI}, for the different prior PDFs we considered. We observe that the adaptive technique is superior to the non-adaptive one. Also, we numerically observe the prior PDFs are different per line of Fig. \ref{fig:Adaptive2} but the MMSE remains the same per line. In other words, no matter which downward path we will choose to follow in Fig. \ref{fig:Adaptive2} the optimal MMSE is the same per line of the tree diagram in Fig. \ref{fig:Adaptive2}. Under this assumption one can generalize the behavior of Figs. \ref{fig:adaptive_PI}, \ref{fig:adaptive_05PI} and \ref{fig:adaptive_01PI} to all downward paths of Fig. \ref{fig:Adaptive2}.

The numerical simulations of this section can be found in \cite{Zhou2023code}.

\section{Conclusions}\label{sec:concl}
In this work we presented a genuine Bayesian phase sensing problem. While the MSE is problematic when the prior PDF has a large variance it is still a valuable tool for narrower prior PDFs. Specifically, we focused on several truncated prior PDFs, which proved to be useful in the context of our general adaptive protocol that takes into account all possible outcomes of the optimal measurement per step, giving rise to the tree diagram of Fig. \ref{fig:Adaptive2}. Our numerical optimization over all possible paths gives the same optimized MMSE (as function of the step $s$) regardless of which downward path we choose in Fig. \ref{fig:Adaptive2}.

Further research ideas on the general subject of the present work include: To examine multi-mode and/or multi-variable systems with any combination of non-classical, Gaussian, and entangled states, aiming to reveal the optimal choices for given sensing tasks. Departing from works (like the current one) that are concerned with fundamentally optimal behavior, it would be interesting to expand the analysis to systems suffering losses and noise. From a more fundamental standpoint it is worthwhile to continue and expand on the direction opened by \cite{rubio2024,Demkowicz2011,Bavaresco2024}, i.e., to consider different cost functions tailored for each sensing problem, even if such studies might rely on numerical evaluations.

The authors acknowledge useful discussions with Jes\'{u}s Rubio (University of Surrey) and Boulat Bash (University of Arizona). B.Z. acknowledges financial support from C.N.G.'s start-up fund (University of Arizona, ECE department). B.Z. and S.G. acknowledge the DARPA IAMBIC Program funded under Contract No. HR00112090128.

\begin{appendices}
\section{Optimal projective measurements}\label{app:Bprime}
\renewcommand{\thesubsection}{\arabic{subsection}}
\def\theequation{A\arabic{equation}}
\setcounter{equation}{0}
\renewcommand{\thefigure}{A\arabic{figure}}    
\setcounter{figure}{0}

Let a Hermitian operator $\hat{K}$, such that
\begin{eqnarray}
    \label{eqB:KGamma0} \hat{K}\hat{\Gamma}_0 = \hat{\Gamma}_0\hat{K}=0.
\end{eqnarray}
We define an operator,
\begin{eqnarray}
    \label{eqB:Bprime} \hat{B}'=\hat{B}+\hat{K},
\end{eqnarray}
where $\hat{B}$ is the operator given in Eq. \eqref{eq:B}.

Let us show that if $\hat{B}$ is a solution to equation $\hat{B}\hat{\Gamma}_0+\hat{\Gamma}_0\hat{B}=2 \hat{\Gamma}_1$, then $\hat{B}'$ is a solution as well.
That means, we need to prove that $\hat{B}'\hat{\Gamma}_0+\hat{\Gamma}_0\hat{B}'=2 \hat{\Gamma}_1$. Starting from the left hand side, we have,
\begin{eqnarray}
   \hat{B}'\hat{\Gamma}_0+\hat{\Gamma}_0\hat{B}' &=&\\
   (\hat{B}+\hat{K})\hat{\Gamma}_0+\hat{\Gamma}_0(\hat{B}+\hat{K})&=&\\
 \label{eqB:proofAppB}  \hat{B}\hat{\Gamma}_0+\hat{\Gamma}_0\hat{B}&=& 2\hat{\Gamma}_1,
\end{eqnarray}
where we have used the condition set in Eq. \eqref{eqB:KGamma0}.

We now want to prove that choosing the solution $\hat{B}'$ (instead of $\hat{B}$), the MMSE defined in Eq. \eqref{eq:delta1} remains the same. We define the quantity,
\begin{eqnarray}
    \label{eqB:deltaprime} \delta'= \text{tr}\hat{\Gamma}_2 - \text{tr}(\hat{B}' \hat{\Gamma}_1).
\end{eqnarray}
From Eqs. \eqref{eqB:Bprime} and \eqref{eqB:deltaprime} we have,
\begin{eqnarray}
\label{eqB:deltaprime2}    \delta'= \delta-\text{tr} (\hat{K}\hat{\Gamma}_1).
\end{eqnarray}
Therefore, we need to prove that $\text{tr} (\hat{K}\hat{\Gamma}_1)=0$. Using Eqs. \eqref{eqB:KGamma0}, \eqref{eqB:proofAppB}, and the cyclic permutation property within a trace, we have,
\begin{eqnarray}
   \text{tr} (\hat{K}\hat{\Gamma}_1) &=& \frac{1}{2} \text{tr}(\hat{K}\hat{B}\hat{\Gamma}_0+\hat{K} \hat{\Gamma}_0 \hat{B})\\
   &=& \frac{1}{2} \text{tr}(\hat{K}\hat{B}\hat{\Gamma}_0) + \frac{1}{2}\text{tr}(\hat{K} \hat{\Gamma}_0 \hat{B})\\
   &=& \frac{1}{2} \text{tr}(\hat{B}\hat{\Gamma}_0 \hat{K}) + \frac{1}{2}\text{tr}(\hat{K} \hat{\Gamma}_0 \hat{B})\\
   &=&0
\end{eqnarray}
and therefore Eq. \eqref{eqB:deltaprime2} gives,
\begin{eqnarray}
    \delta'=\delta.
\end{eqnarray}

\section{The beam splitter generated state and the flat prior PDF}\label{app:BSstate}
\renewcommand{\thesubsection}{\arabic{subsection}}
\def\theequation{B\arabic{equation}}
\setcounter{equation}{0}
\renewcommand{\thefigure}{B\arabic{figure}}
A way to construct a state in the form of Eq. \eqref{eq:GenericState}, is to let the product state consisting of a Fock state and vacuum as input (upper and lower modes respectively) to a beam splitter of transmissivity $\tau$, i.e.,
\begin{eqnarray}
    |\Psi_\tau\rangle = \hat{U}(\tau) |n,0\rangle.
\end{eqnarray}
In this work we define the action of the beam splitter in the Heisenberg picture as,
\begin{eqnarray}
    \begin{pmatrix}
        \hat{b}_1 \\
        \hat{b}_2
    \end{pmatrix}=
    \begin{pmatrix}
        \sqrt{\tau} & \sqrt{1-\tau}\\
        -\sqrt{1-\tau} & \sqrt{\tau}
    \end{pmatrix} \begin{pmatrix}
        \hat{a}_1 \\
        \hat{a}_2
    \end{pmatrix},
\end{eqnarray}
where $\hat{a}_i$ ($\hat{b}_i$) are the input (output) annihilation operators and $i=1,2$ counts the modes starting from the top of Fig. \ref{fig:Phase}. We find,
\begin{eqnarray}
    |\Psi_\tau\rangle = \sum_{l=0}^n c_l(\tau) |l,n-l\rangle,
    \label{eq:BSstate}
\end{eqnarray}
where,
\begin{eqnarray}
    c_l(\tau)=\sqrt{\binom{n}{l} \tau^l (1-\tau)^{n-l}}.
    \label{eq:BScoef}
\end{eqnarray}
Substituting the coefficients of Eq. \eqref{eq:BScoef} into Eq. \eqref{eq:deltaGeneric}, we can obtain the expression of the MMSE for a flat prior PDF and the input state of Eq. \eqref{eq:BSstate}. We note that if the coefficients $c_l(\tau)$ are multiplied by a phase $e^{i \theta_l}$, the MMSE will remain invariant.

We run our numerical optimization (minimization) over $\tau$ of the MMSE of Eq. \eqref{eq:deltaGeneric} using the coefficients of Eq. \eqref{eq:BScoef} (i.e. $a_l=c_l(\tau)$) and for the flat prior PDF. Our constraints are: $0\leq \tau \leq 1$, $\sum_{l=0}^n c_l(\tau)^2=1$, while the mean-photon number constraint comes natural since the state of Eq. \eqref{eq:GenericState} has a fixed photon number equal to $n$. We considered values $n=1,\ldots,100$ and we found the optimal value to be $\tau_{\text{opt}}=0.5$, resulting to the optimal beam splitter generated input state,
\begin{eqnarray}
    |\Psi_{\tau=0.5}\rangle =\left(\frac{1}{\sqrt{2}}\right)^n \sum_{l=0}^n \sqrt{\binom{n}{l} } |l,n-l\rangle.
\end{eqnarray}

\section{State optimization for the flat prior PDF}\label{app:StateOpt}
\renewcommand{\thesubsection}{\arabic{subsection}}
\def\theequation{C\arabic{equation}}
\setcounter{equation}{0}
\renewcommand{\thefigure}{C\arabic{figure}}
\begin{figure}[H]
\centering
\includegraphics[width=0.9\textwidth]{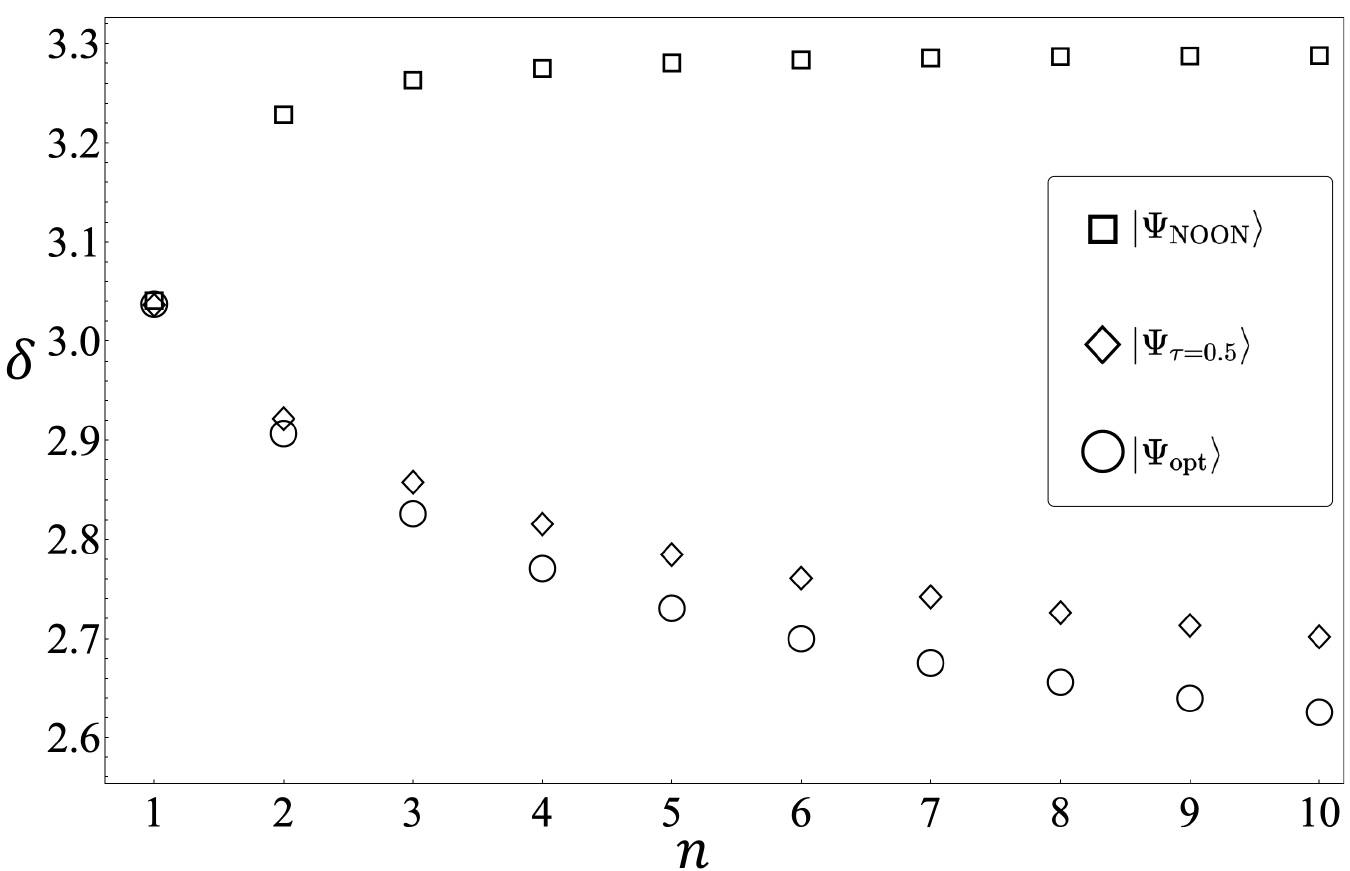}
\caption{The MMSE for the NOON state (squares), the optimized beam splitter generated state (diamonds), and the fully optimized fixed photon number state (circles), as a function of the photon number $n$.}
\label{fig:deltaNOON}
\end{figure}
We numerically find the optimal MMSE for the generic fixed photon number input state of Eq. \eqref{eq:GenericState} for $n=1,\ldots,10$. For said task, since the MMSE depends only on the absolute values of $a_l$, we will assume (only up to the end of this section) that $a_l \geq 0,\ \forall\ l$. Our constraints are: $a_l\geq 0$, $\sum_{l=0}^n a_l^2 = 1$, while, just as in Section \ref{sec:BSstate}, the photon number is fixed as per the form of Eq. \eqref{eq:GenericState}.

In Fig. \ref{fig:deltaNOON} we present said optimized MMSE, and how it compares with the MMSEs of the NOON state and the optimized beam splitter generated state. For the sake of brevity, in Table \ref{tab:OptVals} we give the optimal values of $a_l$ for up to $n=5$, but one can have the numerics run for $n>5$, something that is demonstrated in Fig. \ref{fig:deltaNOON}.
\begin{table}
\caption{The values of the optimal coefficients $a_l^{\text{opt}}$ for values of $n\leq 5$ and the corresponding MMSE $\delta^{\text{opt}}_{n}$. We observe that the coefficients for each $n$ possess a symmetry: The first value is equal to the last value, the second value is equal to the second to last value, etc. Such symmetry is observed in the coefficients of the beam splitter generated state. However, the optimal generic state and the optimal beam splitter generated state are not identical unless $n=1$. A qualitatively similar symmetry has been derived in  \cite{summy1990phase}.}
\label{tab:OptVals}
\begin{tabular}{cccccccc}
\toprule
 $n$ & $\delta^{\text{opt}}_{n}$ & $a_0$ & $a_1$ & $a_2$ & $a_3$ & $a_4$ & $a_5$  \\
\hline
1 & 3.03987 & 0.707107 & 0.707107 & / & / & / & / \\ \hline
2 & 2.90943 & 0.55108 & 0.626595 & 0.55108 & / & / & / \\ \hline
3 & 2.82860 & 0.453382 & 0.542627 & 0.542627 & 0.453382 &/ &/ \\ \hline
4 & 2.77333 & 0.386101 &0.474686 &0.501197 &0.474686 &0.386101 &/ \\ \hline
5 & 2.73304 & 0.336767 &0.420815 &0.457715 &0.457715 &0.420815 &0.336767 \\
\botrule
\end{tabular}
\end{table}
We close this discussion by noting that for $n=1$ the optimal coefficients of the generic state of Eq. \eqref{eq:GenericState} can be found analytically. For said case we find,
\begin{eqnarray}
  \label{eq:deltaoptn=1}  \delta^{\text{opt}}_{n=1} = \frac{\pi^2}{3}-\frac{1}{4}
\end{eqnarray}
while the optimal coefficients (up to an arbitrary phase) are,
\begin{eqnarray}
 \label{eq:OptCoef0}   a_0^{\text{opt}}&=&\frac{1}{\sqrt{2}},\\
  \label{eq:OptCoef1}  a_1^{\text{opt}}&=&\frac{1}{\sqrt{2}},
\end{eqnarray}
which is identical to the NOON state for $n=1$ and to the beam splitter generated state (see Appendix \ref{app:BSstate}) for $n=1$ and $\tau=0.5$.

\section{Adaptive technique with the flat prior PDF}\label{app:adaptiveFlat}
\renewcommand{\thesubsection}{\arabic{subsection}}
\def\theequation{D\arabic{equation}}
\setcounter{equation}{0}
\renewcommand{\thefigure}{D\arabic{figure}}
In this Appendix we use our adaptive technique assuming the flat prior PDF. We do this even knowing that this is a problematic choice for the MSE metric and we draw conclusions on the manifestation of said problematic error metric choice. From Eqs. \eqref{eq:B}, \eqref{eq:flatPrior}, and \eqref{eq:InState_i} we get,
\begin{equation}
\hat{B}_{\text{flat}}=\left(
\begin{array}{cc}
 \pi   & i a_0 a_1^* \\
 -i a_0^* a_1 & \pi   \\
\end{array}
\right).
\label{eq:BiFlat}
\end{equation}
For the flat prior PDF we have shown that the for $n=1$ the optimal state corresponds to $a_0^{(1)}=1/\sqrt{2}=a_1^{(1)}=1/\sqrt{2}$ (see Appendix \ref{app:StateOpt}). Therefore, for said case from Eq. \eqref{eq:BiFlat} we get,
\begin{equation}
\hat{B}_{\text{flat}}=\left(
\begin{array}{cc}
 \pi   & \frac{i}{2} \\
 -\frac{i}{2} & \pi   \\
\end{array}
\right),
\label{eq:B1}
\end{equation}
whose eigenvectors are,
\begin{eqnarray}
    |v_1\rangle&=&\frac{1}{\sqrt{2}}
    \begin{pmatrix}
        i \\ 1
    \end{pmatrix}\\
    |v_2\rangle&=&\frac{1}{\sqrt{2}}
    \begin{pmatrix}
        -i \\ 1
    \end{pmatrix}
\end{eqnarray}
and the corresponding eigenvalues are $v_1=\pi+1/2$, $v_2=\pi-1/2$. The optimal MMSE $\delta^{(1)}$ is given by Eq. \eqref{eq:deltaoptn=1}.

\begin{figure}[H]
\centering
\includegraphics[width=0.9\textwidth]{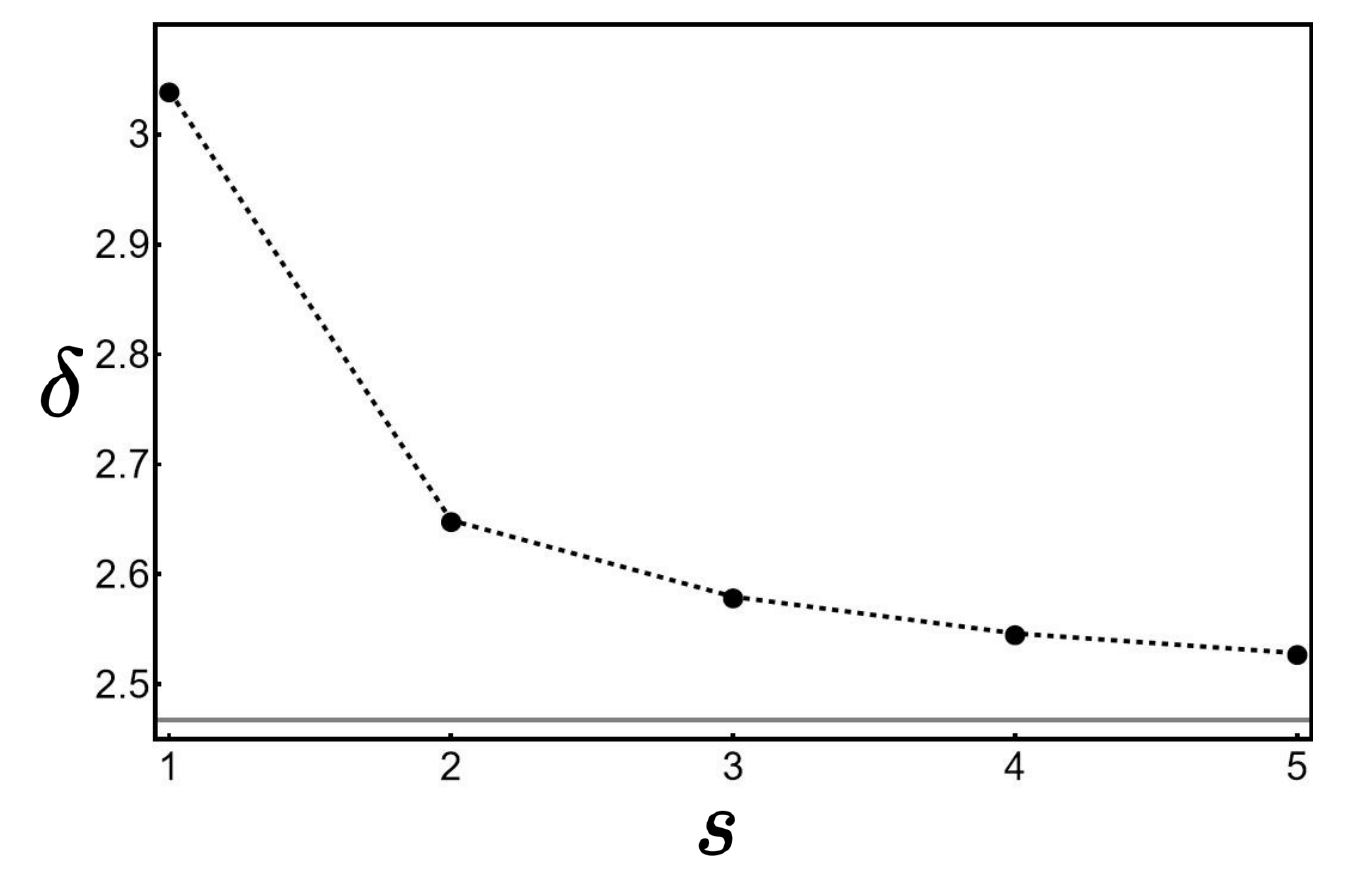}
\caption{The optimized MMSE over the all possible states of Eq. \eqref{eq:InState_i} and all possible prior PDFs for each step $s$. The grey horizontal line corresponds to $\delta^{(\infty)} = \frac{\pi^2}{4}$, i.e., the MMSE for when the prior PDF becomes a double Dirac delta function whose peaks are separated by $\pi$.}
\label{fig:PlotDeltavS}
\end{figure}

In Fig. \ref{fig:PlotDeltavS} we present our findings: The optimal states and measurements for each step, i.e., those corresponding to the $2^{s-1}$ different prior PDFs, are not identical to each other, however all possible outcomes per step lead to the \emph{same} optimized MMSE. As expected, the optimal state can always be written in the dual-rail qubit basis and in particular is always a phase shifted version of the state $|\Psi_1^{\text{opt}}\rangle$ with which we initiated the adaptive protocol.
\begin{figure}[H]
\centering
\includegraphics[width=0.9\textwidth]{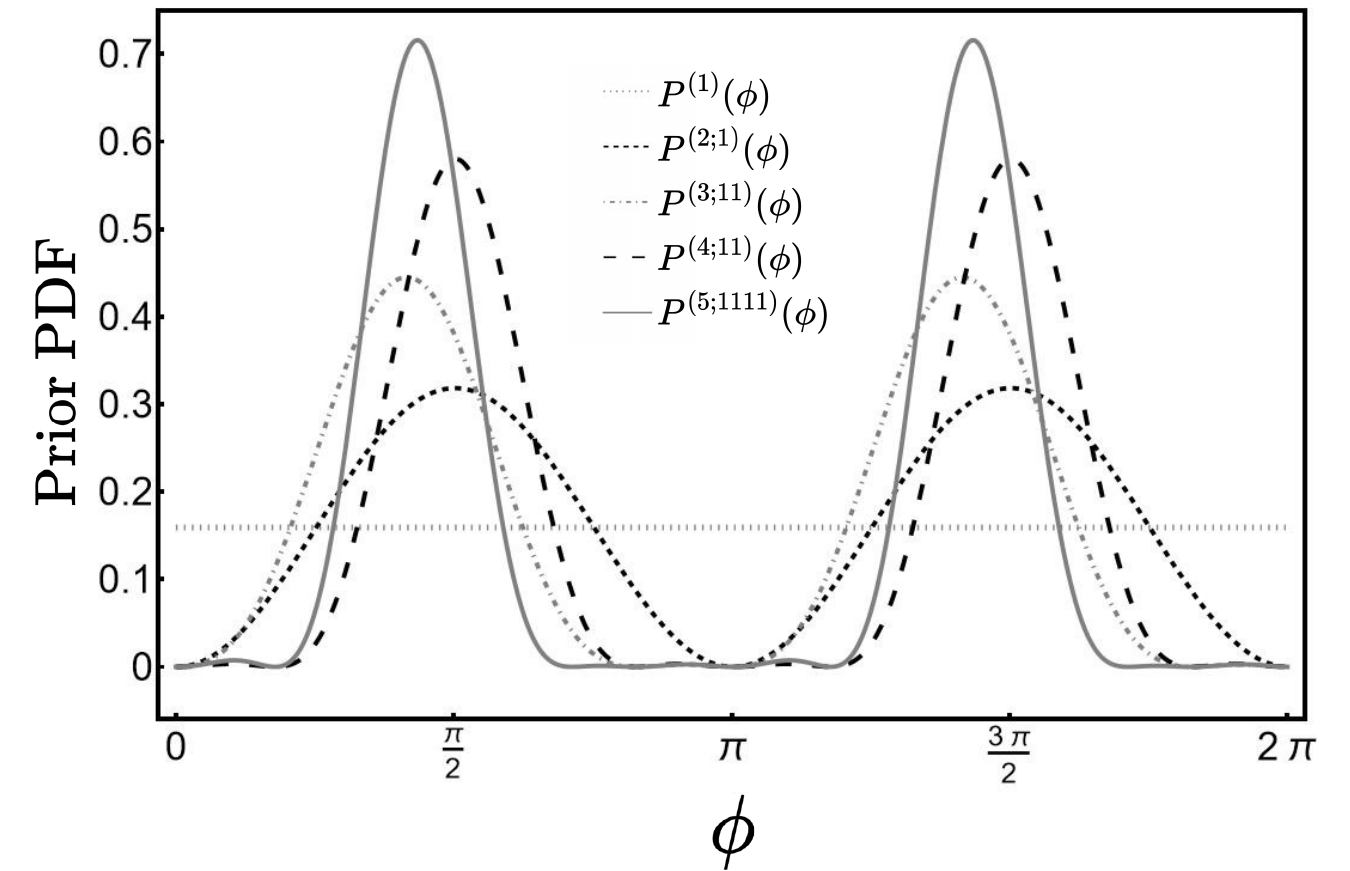}
\caption{The evolution of the prior PDFs for the leftmost path of Fig. \ref{fig:Adaptive2}. The PDFs are plotted as a function the unknown phase $\phi$.}
\label{fig:PriorPlots}
\end{figure}

We demonstrate the evolution of the prior PDF in Fig. \ref{fig:PriorPlots} for a specific path of Fig. \ref{fig:Adaptive2}, namely for the path that corresponds to the largest eigenvalue for each step (there is no physical reason for such choice, however it was computationally convenient to start with). We see that starting from a flat prior, after every step the prior PDF becomes sharper and taller, while the regions equal (or numerically close to) zero become more extended. Different downward paths in Fig. \ref{fig:Adaptive2} give similar results.

Assuming that the optimized MMSE will continue to be the same regardless of the different prior PDF choices, we were able to easily allow $s\gg5$. We see that as $s \rightarrow \infty$ the prior PDF trends to a double Dirac delta function, whose two peaks are separated by $\pi$. For that limiting case, the (analytically calculated) MMSE is,
\begin{eqnarray}
    \delta^{(\infty)} = \frac{\pi^2}{4}.
    \label{eq:deltaInfty}
\end{eqnarray}

\begin{figure}[H]
\centering
\includegraphics[width=0.9\textwidth]{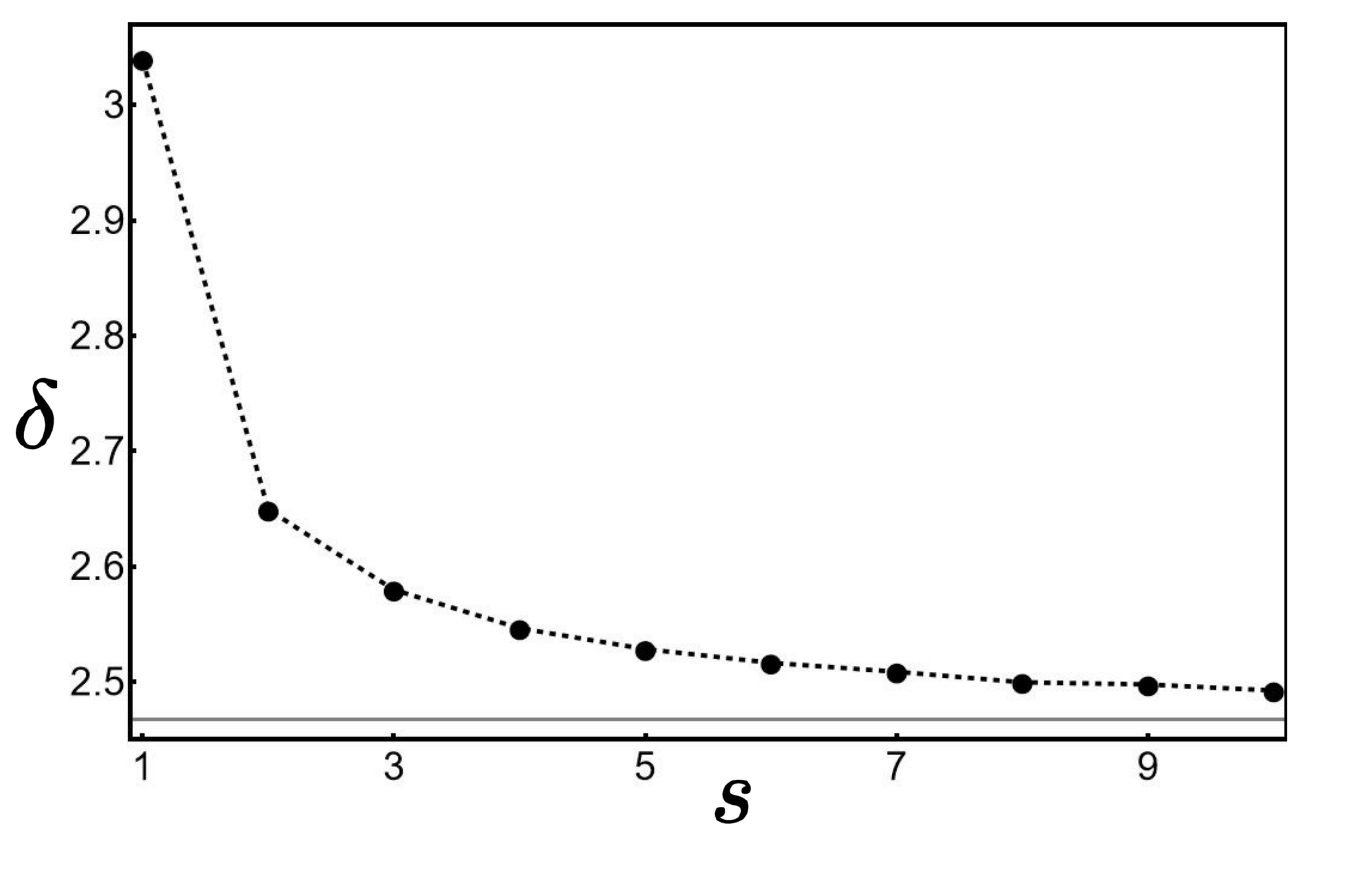}
\caption{The optimized MMSE over the all possible states of Eq. \eqref{eq:InState_i} and the prior PDFs $P^{(s;1\ldots 1)}(\phi)$ for each step $s$. The grey horizontal line corresponds to $\delta^{(\infty)} = \frac{\pi^2}{4}$, i.e., the MMSE for when the prior PDF becomes a double Dirac delta function whose peaks are separated by $\pi$. The values for $s=1,\ldots,5$ are identical to those of Fig. \ref{fig:PlotDeltavS}.}
\label{fig:DeltavS10}
\end{figure}

Under the last assumption, in Fig. \ref{fig:DeltavS10} we show the optimized MMSE per step for up to $s=10$.
\begin{figure}[H]
\centering
\includegraphics[width=0.9\textwidth]{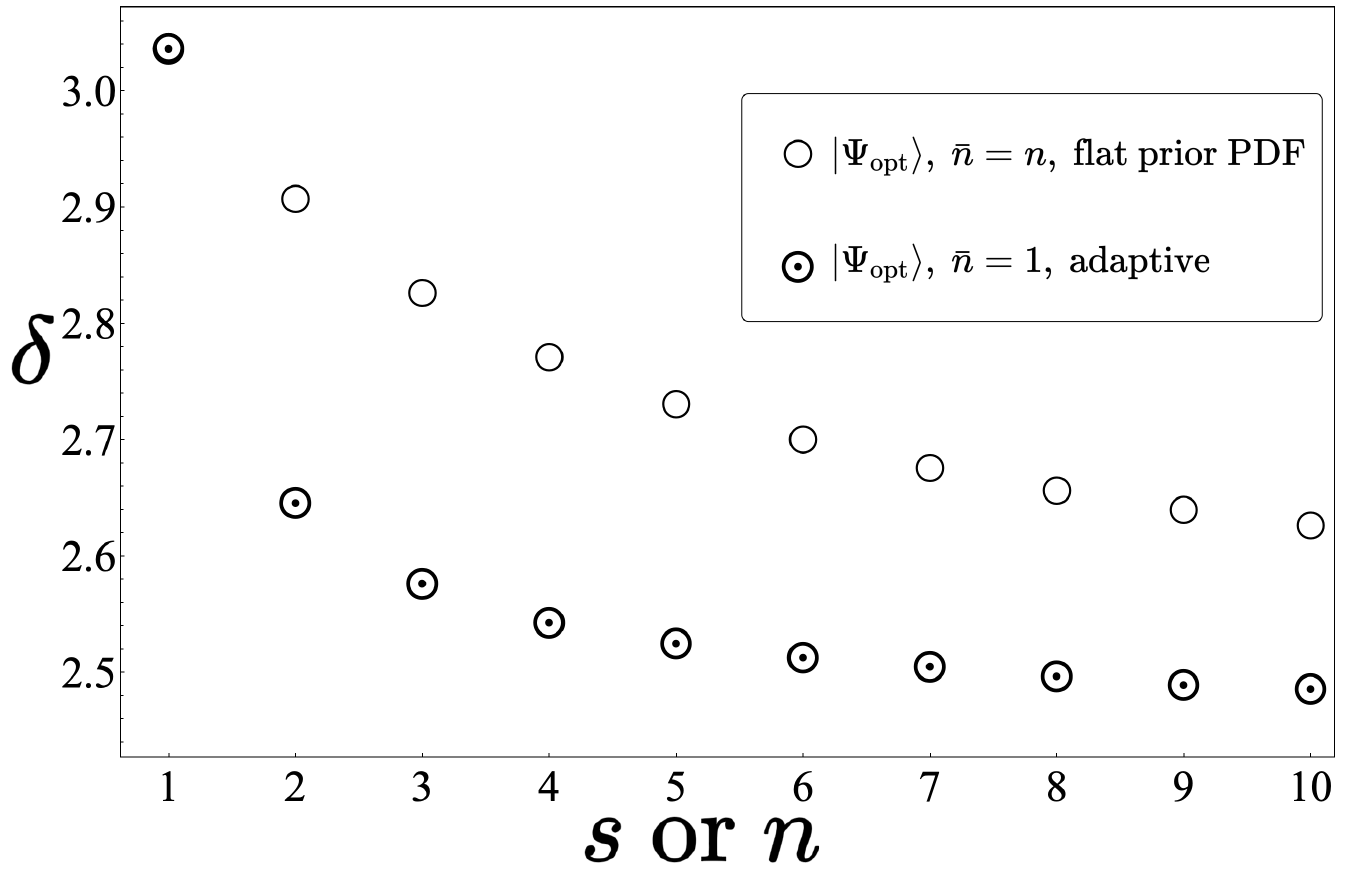}
\caption{The MMSEs of the adaptive technique (circles with dots) and the single-state-use technique (empty circles). For the former the mean photon number of each optimal state per step is $\bar{n}=1$ and at step $s$ we have used a photon budget $\Bar{N}=s \bar{n}=s$. For the latter, the optimal state's mean photon number is $\bar{n}=n$, therefore for this single-step strategy the total mean photon number is $\bar{N}=n$. To compare the two strategies we set $n=s$. The MMSE for the two strategies at $n=s=1$ are numerically identical as expected: For said step $s=1$ we use the flat prior PDF while for the non-adaptive strategy we always use the said prior PDF. The empty circles are identical to those of Fig. \ref{fig:deltaNOON}.}
\label{fig:DeltaVSsn}
\end{figure}

Just like in Section \ref{sec:AdaptiveTrunFlat},  we compare two strategies: (i) The adaptive strategy, and (ii) single-state-use strategy. We show our results in Fig. \ref{fig:DeltaVSsn}, where we see that the MMSE of strategy (i) is less (i.e. better) than the MMSE of strategy (ii) for the some mean photon budget used up to any given step, i.e., the adaptive technique is superior to the non-adaptive one.  Lastly, we note that the Eq. \eqref{eq:deltaInfty} not trending to zero reflects the improper use of the MSE when the initial prior PDF has a large variance.

The numerical simulations of this section can be found in \cite{Zhou2023code}.

\end{appendices}

\bibliography{sn-bibliography}

\end{document}